**Harvesting room-temperature plasticity in ceramics by mechanically seeded dislocations**

Xufei Fang[1,2]*†, Wenjun Lu[3]*†, Jiawen Zhang[3], Christian Minnert[4,5], Junhua Hou[3], Sebastian Bruns[4], Ulrike Kunz[4], Atsutomo Nakamura[6], Karsten Durst[4], Jürgen Rödel[1]*

[1]Ceramics Division, Department of Materials and Earth Sciences, Technical University of Darmstadt, Germany

[2]Institute for Applied Materials, Karlsruhe Institute of Technology, Germany

[3]Department of Mechanical and Energy Engineering, Southern University of Science and Technology, Shenzhen, China

[4]Physical Metallurgy Division, Department of Materials and Earth Sciences, Technical University of Darmstadt, Germany

[5]Empa, Swiss Federal Laboratories for Materials Science and Technology, Thun, Switzerland

[6]Osaka University, Graduate School of Engineering Science, Osaka University, Japan

*Corresponding authors. Email: xufei.fang@kit.edu (XF); luwj@sustech.edu.cn (WL); roedel@ceramics.tu-darmstadt.de (JR)

†These authors contributed equally to this work.

**Abstract:**

The quest for room-temperature ductile ceramics has been repeatedly fueled by hopes for large-scale applications but so far has been not successful. Recent demonstrations of enhanced functional properties in ceramics through judicious dislocation imprint, however, have been sparking renewed interest in dislocation plasticity in brittle ceramics. Here, we propose a facile approach using room-temperature mechanically seeded mobile dislocations with a density of ~$10^{14}$/m$^2$ to significantly improve the room-temperature plasticity of ceramics with a large plastic strain beyond ~30%. The seeded mobile dislocations trigger profuse dislocation multiplication via cross slip and motion. Hence, they offer an avenue to suppress brittle fracture and harvest plasticity in ceramics without any additional high-temperature process. We employ both in situ nano-/micromechanical deformation and ex situ bulk deformation to bridge the length scales. This finding tackles the pressing bottleneck of dislocation engineering in ceramics for achieving ductile ceramics and harvesting both versatile mechanical and functional properties.

**Main Text:**

Ceramic materials have been used by humankind as early as 26,000 BC during the late Paleolithic period. Throughout history it has been accepted that ceramics are hard and brittle, rendering them difficult to deform and prone to fracture. Numerous efforts *(1-4)* have been made to understand the brittleness of ceramic materials from a fundamental academic aspect, as well as to achieve ductile and deformable ceramics with improved performance for versatile engineering applications.

Dislocations, one-dimensional line defects and the main carriers of plastic deformation in crystalline solids, more widely known in metals, are one most promising candidate for achieving ductile ceramics. Extensive studies were conducted in the past on bulk single crystals such as LiF *(5)*, NaCl *(6)*, KCl and MgO *(7)*, predominantly with rock salt structure. However, even for such plastically deformable ceramics, a fundamental challenge remains concerning crack formation and fracture after merely a few percent of plastic strain, as discussed by Stokes et al. in 1958 *(8)*, followed by Argon & Orowan for the case of MgO crystals in the early 1960s *(9, 10)*. Up to date, there is still a lack of both fundamental understanding and an effective method in suppressing crack formation while promoting large plasticity in ceramics across the length scales.

Meanwhile, the emerging dislocation-based functionality in ceramic single crystals *(11-13)* is reigniting the dislocation mechanics study in ceramics *(3, 14)*. Contrasting the conventional picture that dislocations are undesirable in semiconductors, dislocations in ceramic oxides are being engineered to harvest versatile physical properties including piezoelectricity *(13)*, superconductivity *(15)*, electrical conductivity *(16)*, photoconductivity *(17)*, thermal conductivity *(18)*, as well as mechanical properties such as enhanced ductility *(19)* and fracture toughness *(20)*. Dislocations may hold the key to a new horizon of *dislocation technology (11, 12)* for advanced ceramic engineering. *The pressing bottleneck, however, lies in the lack of efficient dislocation engineering into brittle ceramics.* To address this limitation, recent research endeavors have been focusing on mechanical deformation *(14, 21)*, bicrystal fabrication *(22)*, flash sintering *(19)*, thin film growth *(18)*, and irradiation *(23)*, among which the pivoting endeavor focuses on mechanical deformation of ceramic crystals.

Room-temperature deformation of ceramics has enjoyed a long history in bulk deformation *(5, 21)*. Nowadays, the mainstream deformation approach for ceramics is small-scale testing such as nano-/micropillar compression *(24)* and nanoindentation tests *(25)*. The small deformed volumes (in nano-/micropillar compression) minimize the flaw population and favor plastic flow over cracking or suppress crack propagation by means of locally high compressive hydrostatic stress (in nanoindentation). The size effect in metals, "smaller is stronger" *(26, 27)*, reveals a widely accepted physical picture where smaller samples exhibit much higher yield strength with no (or few) dislocations prior to deformation, dominated by dislocation nucleation. In contrast, the yield strength decreases with increasing deformation volume and number of dislocations therein, owning to dislocation multiplication and motion (**Fig. 1A-D**). In bulk ceramics, conventional sintering renders a much lower density of grown-in dislocations (e.g., ~$10^{10}$/m$^2$, see Supplementary **Fig. S1**, and can be regarded as dislocation-free or dislocation-scarce). Concerning plasticity in ceramics, this shifts the bottleneck to dislocation nucleation before cracking sets in (**Fig. 1E-G**). However, homogeneous dislocation nucleation is energetically unfavorable and requires stresses approaching the theoretical shear strength of ~$G/2\pi$ (*G* is the shear modulus), which is about 1-2 orders of magnitude higher than the actual fracture strength of most ceramics.

Hence, we are faced with the dilemma of circumventing dislocation nucleation while promoting dislocation motion and multiplication. Our approach lies in directionally imprinting a high density of mobile dislocations from the surface at room temperature. These provide an originally confined small plastic zone, which permeates throughout the whole bulk during successive loading. With this treatment, one shall expect plastic yield of ceramics via engineered existing mobile dislocations to achieve appreciable plastic deformability, a similar effect as in metals. This concept is visualized in **Fig. 1H-J**, where we inject mechanically seeded dislocations (details in Supplementary Materials & Methods, **Fig. S2-3** and **Movie S1**). We demonstrate that this approach effectively suppresses crack formation and significantly promotes dislocation plasticity across the length scales, ranging from nanoscale to bulk testing.

We start with nano- and micropillar compression (Supplementary Materials & Methods), followed by mesoscale and bulk tests. The *in situ* nanopillar compression in the TEM (**Fig. 2**) demonstrates that the dislocation-free pillar (**Fig. 2A-C**) suffers from abrupt fracture during compression (Supplementary **Movie S2**). The fracture surface is on the {011} plane, in line with previous reports that {011} planes form low-energy fracture planes in SrTiO$_3$ *(20, 28)*. In contrast, the nanopillar with seeded dislocations (**Fig. 2D-F**) was captured with profuse multiplication and motion of dislocations (Supplementary **Movie S3**), without cracking even up to ~32% plastic strain. Note that before loading, the seeded dislocations (with a density of ~$10^{14}$/m$^2$, black line contrasts in **Fig. 2D-F**) appear randomly distributed, which is a signature of the surface grinding and polishing procedure for introducing the seeded dislocations. Later during mechanical deformation, dislocations were aligned (indicated by the 45° inclined line features in **Fig. 2E, F**) on the {110} slip planes activated at room temperature for SrTiO$_3$ *(29)*. The deformation processes are schematically illustrated in **Fig. 2G, H**.

With an increased pillar size up to ~4 μm in diameter (~8 μm in height), we observed a consistent deformation behavior (**Fig. 3**): the dislocation-free micropillar shattered (**Fig. 3C**) at an elastic limit up to ~4.5 GPa (**Fig. 3A**). In comparison, the micropillar with seeded dislocations deformed plastically up to a plastic strain of $\varepsilon_p$ ~25% (**Fig. 3B**), exhibiting a metal-like flow behavior with a yield stress lower than 500 MPa and multiple slip traces on the pillar surface. A clear work hardening behavior in the dislocation-seeded pillars was also observed, which was not commonly seen in the deformation of ceramics. Compared to the pillar with $\varepsilon_p$ ~8% (**Fig. 3D**), the slip traces in the pillar with $\varepsilon_p$ ~25% (**Fig. 3E**) become so fine that they are hardly distinguishable. Note that tests on the two dislocation-seeded micropillars (**Fig. 3D, E**) were both interrupted at designated strains before fracture for post-mortem TEM analysis to examine the dislocation structure inside the pillars. The mechanically seeded dislocations after surface grinding and polishing were confined to the highlighted top region (blue rectangular boxes, see also **Fig. S2-3**, with ~2 μm in depth and much shallower than the micropillar height). These two TEM "snapshots" (**Fig. 3F, H**) visualize the seeded dislocations in the near-surface region glided, multiplied, and penetrated into the lower region of the micropillars. The green dashed arrow (**Fig. 3F**) indicates the front of advancing plasticity during pillar deformation, providing strong evidence of dislocation multiplication via cross slip. This cross slip behavior agrees with experimental evidence from dark-field X-ray diffraction and molecular simulation on SrTiO$_3$ *(20)*. The gradient in dislocation density (**Fig. 3G, I**) along the loading direction of the micropillars can be readily quantified with up to ~$10^{15}$/m$^2$ near the pillar surface and a reduction of dislocation density of about a factor of 10-100 approaching the bottom. The gradient in dislocation density directly after seeding is also reflected in Supplementary Materials (**Fig. S2-3**).

Several micropillars with/without seeded dislocations were tested for reproducibility (**Fig. S4**). An overview of the stress-strain curves for the deformed micropillars are summarized in **Fig. S5**. On average, the fracture stress for dislocation-scarce pillars approaches 3.1 GPa (**Fig. S6**). In contrast, a closer examination of the stress-strain curves for the dislocation-seeded micropillars reveals a yield strength of ~430 MPa (indicated using the 0.2% strain offset, see Supplementary **Fig. S5B**). The taper in the micropillars (**Fig. 3**) may favor the stabilization of the pillar geometry, however, it is not expected to dominate the plastic deformation as the previous *in situ* TEM nanopillar (**Fig. 2D-F**) exhibits no taper.

The consistency in the deformation behavior for nanopillars and micropillars is excellent, but the true challenge lies in scaling up the sample size and deformation volume, which is of paramount relevance for mass production and large-scale exploitation of dislocation-tuned functionalities *(14)*. It remains long debated whether or not the observations and mechanisms derived from *in situ* TEM and small-scale deformation can be compared and transferred to bulk behavior *(30)*. To address this pertinent issue, we further performed mesoscale spherical indentation and scratching tests in cyclic deformation. Although using a spherical indenter (with a diameter of 2.5 mm) induces changes in the stress state compared to the uniaxial pillar compression, the shear and compressive stresses still dominate. As for the choice of using cyclic deformation, we particularly performed repeated indentation/scratching on the same location on the sample surface, using the dislocations generated in the 1$^{st}$-cycle deformation as mechanically seeded dislocations to promote further dislocation multiplication and motion. The results (**Fig. S7-8**) prove that a characteristic plastic zone size of ~100 μm and larger at room temperature can be achieved, with profuse dislocation multiplication from the mechanically seeded dislocations. This cyclic approach assists to mitigate possible artefacts and complexity caused by the increased deformation size: (i) a larger indenter can probe, with a higher chance, existing dislocations (including grown-in dislocations) for samples even without mechanically seeded dislocations (Supplementary Sec.1.2C for estimation); (ii) the sample surface roughness may serve as source for heterogeneous dislocation nucleation *(31)*; (iii) the imperfect indenter surface (with extrusions or roughness) induces local stress concentration for facile dislocation nucleation.

In a perfect crystal without flaws, plasticity starts with homogeneous dislocation nucleation, which is energetically unfavorable for both metals and ceramics at room temperature, requiring an extremely high shear stress approaching the theoretical limit ~$G/2\pi$ (e.g., ~5.2 GPa for Ni *(32)*, ~5 GPa for Cu *(33)*, ~16 GPa for Mo *(34)*, ~22 GPa for W *(35)*, and ~17 GPa for SrTiO$_3$ *(25)*). Nevertheless, plasticity prevails over fracture in perfect or defect-scarce crystals due to the fact the theoretical shear strength (~$G/2\pi$) is lower than the theoretical fracture strength (~$E/10$, $E$ being Young's modulus) *(25, 36)*. This rationalizes the plasticity favored by nanoindentation pop-in with a very sharp tip (with a tip radius in the submicron range) *(25)* and the plastic deformation of brittle materials during nano-/micropillar compression, as evidenced on, but not limited to, carbides *(37)*, nitrides *(4)*, halide perovskites *(38)*, Si *(39)*, and diamond *(40)*, most of which are considered brittle at room temperature. This phenomenon is coined size-dependent brittle to ductile transition (BDT) *(41, 42)*. Extensive plasticity in small ceramic samples may be attributed to: (i) the minimization of existing flaws/defects in the pillars, hence a much higher shear stress can be sustained to facilitate dislocation nucleation before fracture sets in; (ii) the free surfaces of the pillars and the large surface-to-volume ratio can facilitate dislocation escape to the surface to reduce the chance of internal cracking due to intersecting slip planes *(41)* or cracking induced by dislocation pileup *(28)*. In contrast, in bulk deformation the chances are much higher for the activated slip planes to intersect and for dislocations to interact and lock to become immobile and act as stress concentration sites for cracks to initiate and propagate inside the crystal bulk, which was evidently demonstrated by Argon & Orowan *(10)* in MgO bulk compression. This may also explain the gap between the maximum plastic strain in our micropillar tests and the bulk compression on SrTiO$_3$ (Supplementary **Table S1**).

Unlike the size effect in metals ("smaller is stronger" *(26, 27)*) that addresses the size-dependent plastic yield strength, here in ceramics the size effect needs to be understood via the competition between cracking and plasticity *(41, 43)*, namely "smaller is ductile" (see compiled data on ceramics in **Fig. 4A**). Concerns have been raised regarding the critical sample size for BDT (noted as $L_{BDT}$) particularly in nano-/micropillar compression *(41, 44, 45)*. The compendium of literature data in **Fig. 4A** demonstrates that a size transition occurs in the micrometer regime with plasticity ranging up to about 20%. Our work clearly demonstrates that SrTiO$_3$ with mechanically seeded dislocations effectively pushes the upper bound of $L_{BDT}$ to a much higher value (~4 μm) for pillar compression, contrasting to a previously reported very small pillar diameter (below ~180 nm) *(44)* (**Fig. 4A**). Simultaneously, the plastic strain reaches much higher values than in the literature on SrTiO$_3$.

Due to the varying volume bridging the length scales, it is not practical to use the term *dislocation density* here, as it is a volume-scaling invariant parameter. Instead, we use the spacing between the dislocations ($1/\sqrt{\rho}$, $\rho$ is the dislocation density) compared with the sample's characteristic length $D$ (e.g., diameter). With $1/\sqrt{\rho} > D$, the plasticity is dominated by dislocation nucleation, while $1/\sqrt{\rho} < D$ points to dislocation multiplication and motion (as in the case of mechanically seeded dislocations). The schematic in **Fig. 4B** illustrates the yield strength of dislocation-mediated plastically deformable ceramics as a function of $D\sqrt{\rho}$, where the sample volume containing a higher number of directionally seeded dislocations is more likely to plastically deform. However, with dislocation density beyond some critical value $N_c$ (namely, $D\sqrt{\rho} \gg N_c$), strong dislocation work hardening occurs (**Fig. 3B**), and further dislocation pileup may induce crack formation.

For dislocation-tuned functionality, one critical requirement is the high dislocation density *(46, 47)* in a sufficiently large volume, which can be achieved most effectively via dislocation multiplication. Our approach here proves to be a viable strategy using mechanically seeded dislocations. We emphasize that a small plastic zone suffices, capable of penetrating the whole bulk. Hence, directional, simple surface imprint is accessible for complex-shaped volumes for future wide-scale implementation. In the current case, the dislocation multiplication mechanisms at room temperature such as Frank-Read sources and cross slip mechanisms are expected to operate *(48)*. The post-mortem snapshots of the dislocation features in the ~8% and ~25% pillars (**Fig. 3F, H**) as well as video (Supplementary **Movie S3**) captured for the nanopillars deformed via *in situ* TEM provide direct evidence for cross slip. For cross slip to occur, screw dislocations are required *(48)*. The detailed analysis of the Burgers and line vectors of the seeded dislocations (**Fig. S9-11**) identified a major fraction of the dislocations to be mixed type, consistent with the observation by Jin et al. *(49)* and Matsunaga et al. *(50)*. Furthermore, the dislocation core is expected to be compact without dissociation or having a high stacking fault energy (SFE) with much smaller SF width to facilitate cross slip. The room-temperature dislocation core structure in SrTiO$_3$ has been extensively discussed, which is believed to have a glide dissociation configuration *(51, 52)*. The SFE for SrTiO$_3$ was reported to be $136 \pm 15$ mJ/m$^2$ at room temperature *(50)*, which is rather high and leads to a narrow SF width (several nm *(50)*) between the two colinear partials. Our direct observation of significant cross slip events in SrTiO$_3$ (**Fig. 2-3**) facilitates the understanding of the room-temperature bulk deformation of this model material, where slip band widening is frequently observed.

Last but not least, we discuss the yield strength as function of dislocation density by fully accounting for the nanopillars and micropillars tested here as well as the bulk compression tests with available literature data (Supplementary **Table S1**). To streamline the discussion, the data compiled here are exclusively on SrTiO$_3$ crystals tested on (001) surface or along <001> direction. The yield strength follows the general trend observed in metals *(53)* (see **Fig. S12A**), namely, first decreases and then increases as a function of the dislocation density. In the regime where the dislocation density is extremely high (~$10^{14}$/m$^2$), the increased yield strength (**Fig. S12B**) is likely due to dislocation hardening.

The term *existing dislocations* requires attention. One example is the grown-in dislocations, which can be immobile at room temperature if they do not belong to the slip systems required for room-temperature deformation. Effective existing dislocations need to be *mobile* as in the case of mechanically seeded dislocations via grinding, polishing, indentation, or scratching at room temperature. These mechanically-induced dislocations belong to the room-temperature slip systems by nature, hence are effective sources for subsequent plastic deformation. An important characteristic for these materials (e.g., SrTiO$_3$, MgO, ZnS, and CaF$_2$) lies in good dislocation mobility with low lattice friction stress at room temperature. This is evidenced by their low yield strength in single-crystal bulk compression tests, for instance, ~120 MPa for SrTiO$_3$ *(54)*, ~20 MPa for MgO *(55)*, ~60 MPa for ZnS under white light (in darkness it is ~30 MPa) *(3)*, and ~15 MPa for CaF$_2$ *(56)*, all of which are close by the very ductile FCC metals summarized in *(57)*. The concept of mechanically seeded dislocations for room-temperature plasticity can well be extended to other technically important oxides that have an intermediate brittle-to-ductile transition temperature (BDTT) such as TiO$_2$ (BDTT around 600 °C *(58)*), this is in line with the flash-sintered TiO$_2$ *(19)* as well as high-temperature pre-deformed TiO$_2$ *(59)* that both exhibit ~10% plastic strain at room temperature in micropillar compression, owing to the enrichment of existing dislocations and stacking faults. However, challenge remains for extending to room-temperature bulk plastic deformation for such ceramics as TiO$_2$ because the yield strength are still very high (~4 GPa) *(59)*, likely owing to the intrinsic high lattice friction stress, hence poor room-temperature mobility, in such ceramics.

In summary, contrasting the conventional brittle picture, some ceramics appear to share fundamentally similar dislocation mechanisms to a large extent with metals during plastic deformation. If we contrast ceramics and metals which both reveal low lattice friction stress, then ceramics need to be furnished with a high density of mobile dislocations as additional step, for large plastic deformation to occur. This design concept significantly improves the capability of ceramics to plastically deform at room temperature without any additional high-temperature process, as demonstrated on the model system SrTiO$_3$ crystal that exhibits a record-breaking plastic strain up to ~32% without fracture at sub-micrometer scale and ~25% at microscale. The predominating mixed-type dislocations are found to play a key role in promoting cross slip for dislocation multiplication. As the number of existing mobile dislocations in the deformed volume is the controlling factor, the seemingly contradictory results reported in literature regarding the size-dependent competition between cracking and plastic deformation in ceramics across the length scale is clarified. This approach holds great potential for general applicability in various ceramic crystals and will pave the road for the new research trajectory of dislocation-tuned functional ceramics by providing the platform of room-temperature dislocation engineering.

**Acknowledgments:** The authors acknowledge the use of the facilities at the Southern University of Science and Technology Core Research Facility. We also thank the helpful discussions with L. Porz, A. Klomp, K. Albe, S. Zhang, and S. Ogata.

**Funding:** this work is supported by the European Union (ERC Starting Grant, Project MECERDIS, grant no. 101076167). Views and opinions expressed are however those of the authors only and do not necessarily reflect those of the European Union or the European Research Council. Neither the European Union nor the granting authority can be held responsible for them. J. R. is supported by German Research Foundation (DFG) (grant no. 414179371). W. Lu is supported by Shenzhen Science and Technology Program (grant no. JCYJ20230807093416034), National Natural Science Foundation of China (grant no. 52371110) and Guangdong Basic and Applied Basic Research Foundation (grant no. 2023A1515011510).

**Author contributions:** Conceptualization: XF; Methodology: XF, WL, CM, SB, KD, JZ, JH, UK, AN; Investigation: XF, WL, CM, JZ, AN; Visualization: WL, XF; Funding acquisition: XF, WL, JR; Project administration: XF, WL, JR; Supervision: XF, WL, JR; Writing – original draft: XF; Writing – review & editing: WL, ZJ, CM, SB, UK, JH, AN, KD, JR

**Competing interests:** Authors declare that they have no competing interests.

**Data and materials availability:** All data are available in the main text or the supplementary materials.

**Supplementary Materials**

Materials and Methods

Supplementary Text

Figs. S1 to S12

Tables S1

References (*60-78*)

**Fig. 1. Illustration of the design concept for achieving exceptional room-temperature ductility in ceramics.** (**A**): Conventional deformation approach for pristine metal/ceramic crystals with low density of existing/grown-in dislocations (in red); (**B-D**) widely accepted size effect "smaller is stronger" in metal deformation: (**B**) smaller deformation volume without dislocations requires dislocation nucleation and gives higher yield strength, (**C**) large deformation volume with existing dislocations displays lower yield strength via dislocation multiplication and motion; (**E-G**) room-temperature, fracture-dominated deformation of ceramics in the conventional sense: (**E**) smaller deformation volume without existing defects (flaws, dislocations) can deform plastically to a limited plastic strain, (**F**) larger deformation volume often leads to brittle fracture due to crack propagation. (**H-J**) Conceptualized new deformation strategy with mechanically seeded dislocations for ceramic crystals: (**H**) mechanically seeded dislocations are inserted from the sample surface; (**I**) expected large plastic deformation in ceramic samples with seeded dislocations; (**J**) anticipated much improved ductile behavior in the illustrative stress-strain curves, depicting also dislocation-based size effect in ceramics after mechanical seeding in ceramics at room temperature. Details on the mechanical dislocation seeding are given in *Supplementary Sec. 1 Materials & Methods.*

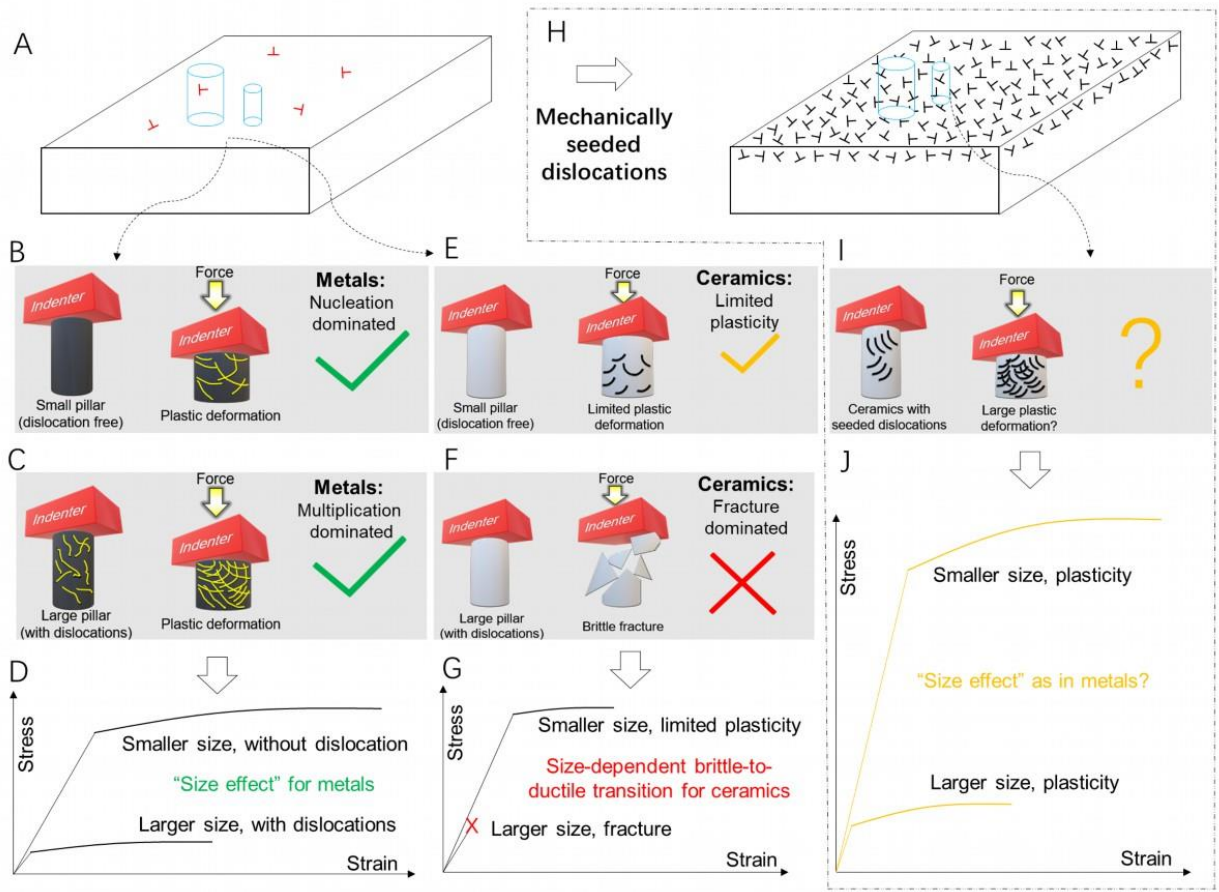

**Fig. 2. In situ nanopillar compression via STEM (scanning transmission electron microscopy):** (**A**) Dislocation-free SrTiO$_3$ nanopillar before compression, with compression direction of <100>, and TEM zone axis of <001>. (**B, C**) Dislocation-free nanopillar exhibits shear-induced brittle fracture along the {110} plane during compression. (**D, E, F**) Nanopillar with mechanically seeded dislocations under different compression strain (0, ~16%, and ~32%) The color bar in (**D**): red indicates a higher dislocation density in the top region of the nanopillar and blue a lower dislocation density at the bottom. The dislocation density gradient originates from the mechanical seeding process (in **Fig. S2-3**). (**G, H**) Schematic illustration of the brittle fracture of the dislocation-free nanopillar during compression, and large plastic deformation of the nanopillar with seeded dislocations.

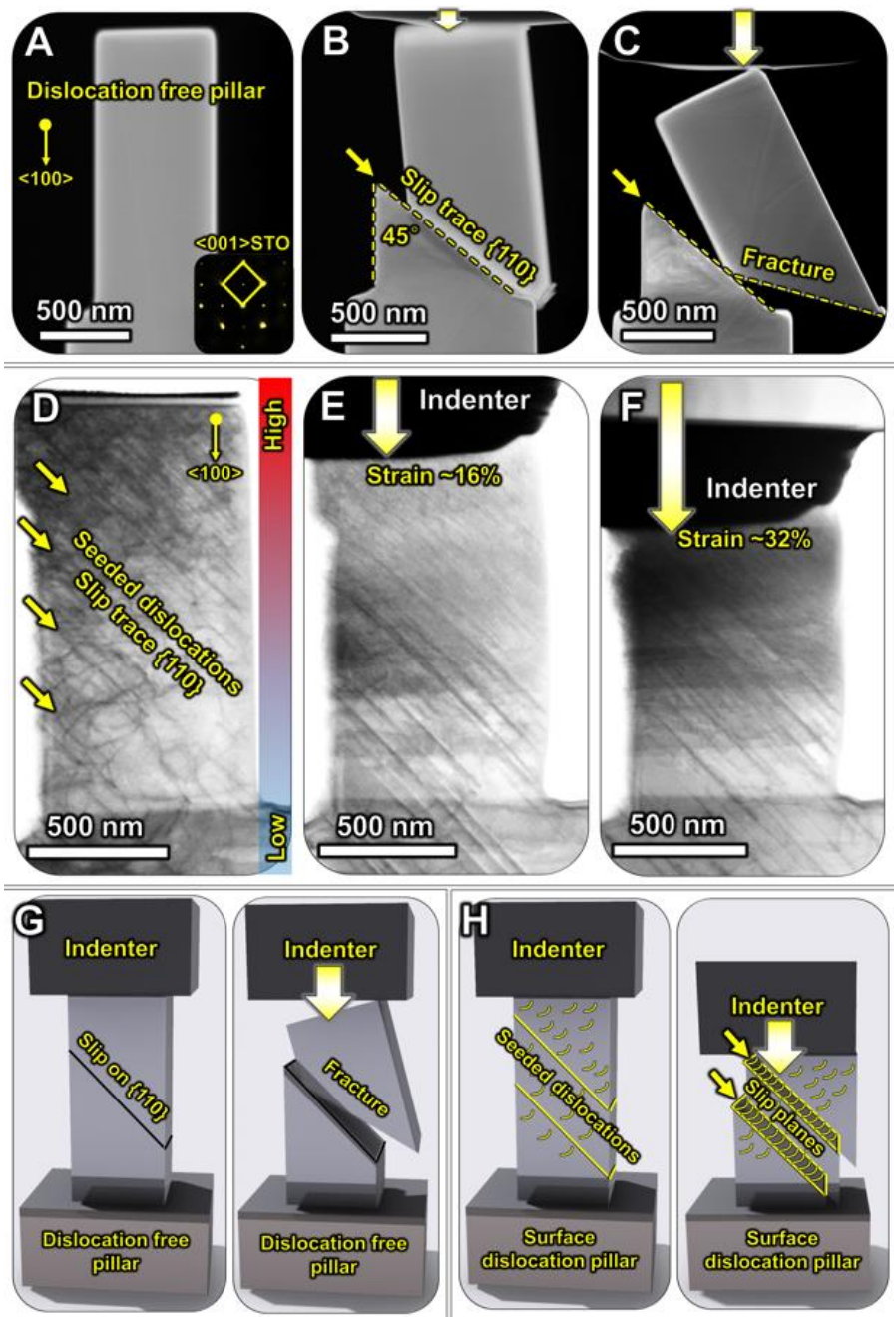

**Fig. 3. Representative deformation results of the micropillar compression**: (**A**) Stress-strain curve for dislocation-free SrTiO$_3$ micropillar suggests an abrupt fracture at the elastic limit. (**B**) Micropillar with seeded dislocations displays compression strain up to ~8% and ~25%, with both tests interrupted before pillar fracture. (**C**) Dislocation-free micropillar after compression features brittle fracture; (**D, E**) Micropillar with seeded dislocations after ~8% and ~25% strain, corresponding to (**B**). (**F**) ADF (annular dark-field)-STEM image reveals the dislocation structure of a micropillar with seeded dislocations with ~8% strain, and (**G**) the statistical depth-dependent dislocation density for the sample in (**F**). (**H**) ADF-STEM image depicts the dislocation structures of micropillar with seeded dislocations with ~25% strain and (**I**) the statistical depth-dependent dislocations for the sample in (**H**).

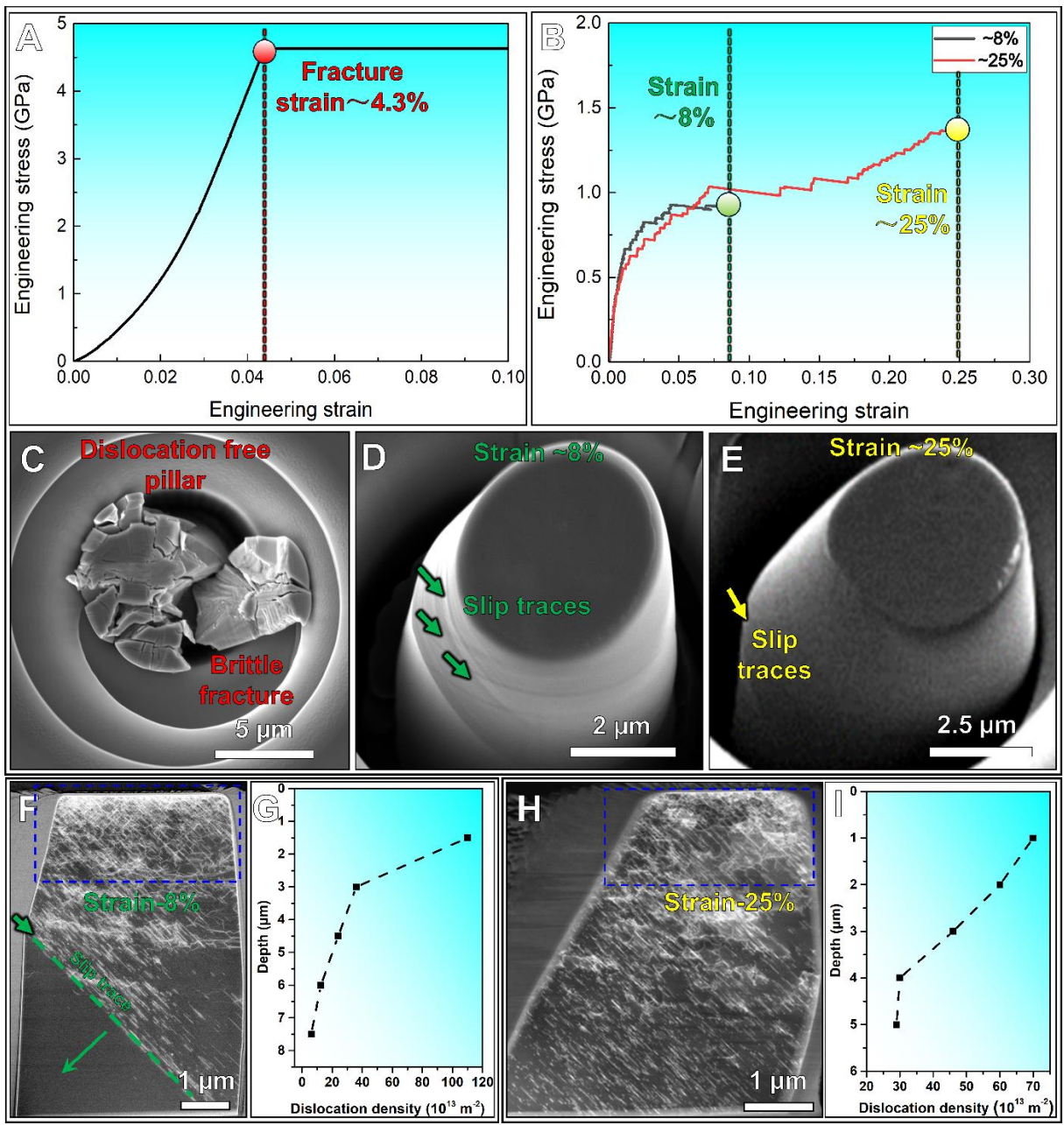

**Fig. 4. Size effect and impact of existing dislocations on plastic deformation of ceramics:** (**A**) Compilation of pillar compression tests for various brittle materials, with a focus on plastic strain as a function of the size of the pillar, indicating a "smaller is ductile" trend. The current work is indicated by the green stars. The reference list for extracting these data is given in *Supplementary Text*. (**B**) Yield strength as function of the unified parameter $D$ vs. $1/\sqrt{\rho}$: with increasing number of mobile dislocations, the yield strength decreases and the chance of plastically deforming ceramics increases, but too high a density leads to dislocation work hardening, and may also result in further dislocation pileup and cracking. $D$ is the characteristic length of the deformation volume, $\rho$ is dislocation density, $N_c$ indicates a critical number of dislocations in the deformed volume.

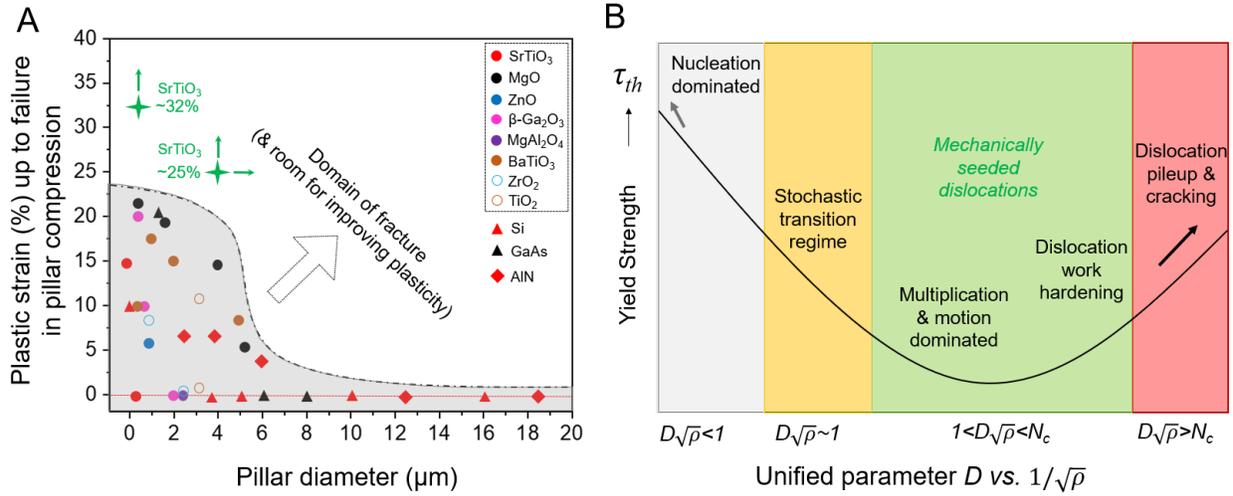

*Supplementary Materials*
*for*

**Harvesting room-temperature plasticity in ceramics by mechanically seeded dislocations**

Xufei Fang[1,2]*†, Wenjun Lu[3]*†, Jiawen Zhang[3], Christian Minnert[4,5], Junhua Hou[3], Sebastian Bruns[4], Ulrike Kunz[4], Atsutomo Nakamura[6], Karsten Durst[4], Jürgen Rödel[1]*

[1]Ceramics Division, Department of Materials and Earth Sciences, Technical University of Darmstadt, Germany

[2]Institute for Applied Materials, Karlsruhe Institute of Technology, Germany

[3]Department of Mechanical and Energy Engineering, Southern University of Science and Technology, Shenzhen, China

[4]Physical Metallurgy Division, Department of Materials and Earth Sciences, Technical University of Darmstadt, Germany

[5]Empa, Swiss Federal Laboratories for Materials Science and Technology, Thun, Switzerland

[6]Osaka University, Graduate School of Engineering Science, Osaka University, Japan

*Corresponding authors. Email: xufei.fang@kit.edu (XF); luwj@sustech.edu.cn (WL); roedel@ceramics.tu-darmstadt.de (JR)

†These authors contributed equally to this work.

## Sec.1. Materials and Methods

### S1.1. Material selection

Single-crystal $SrTiO_3$ (grown via Verneuil technique, Alineason Materials Technology GmbH, Frankfurt am Main, Germany) was primarily used for the mechanical testing at room temperature. The choice of ceramic single crystals eliminates the complexity of grain boundaries, pores, and pre-existing flaws as in the case of polycrystalline samples, allowing us to focus on the fundamental dislocation mechanisms. This is because ceramics at room temperature feature a limited number of available slip systems (normally only 2 independent slip systems at room temperature (*60*), not fulfilling the required 5 independent slip systems for general plastic deformation in polycrystalline samples according to the von Mises or Taylor criterion (*60*)). The limited available number of slip systems in ceramics at room temperature does not favor slip transmission at grain boundaries and can easily lead to dislocation pileup and crack formation. Note that single-crystal $SrTiO_3$ exhibits room-temperature bulk plasticity (*29, 54, 61*) in mm-sized samples but was reported to be brittle at small scale unless tested in pillar compression with a diameter smaller than ~180 nm (*44*). This allows us to validate the effectiveness of the mechanically seeded dislocations for the enhanced plasticity and crack suppression to bridge this huge gap across the length scale.

Unless specified, the single-crystal $SrTiO_3$ samples used are tested in <100> direction. The sample surfaces were prepared by careful polishing to a surface roughness of <1 nm as received. To ensure that no additional surface dislocations were induced during sample preparation, we performed chemical etching to reveal dislocation etch pits prior to any mechanical deformation. The chemical etchant used was made of 15 mL 50% $HNO_3$ with 16 drops of 50% HF and applied for ~20 s to reveal the dislocations intersecting the surface. Representative ceramic oxides including single-crystal $SrTiO_3$ (001) surface with extremely low pre-existing dislocation density (~$10^{10}/m^2$) are presented later in Supplementary **Fig. S1**. These samples with extremely low dislocation density are referred to as reference or pristine samples.

### S1.2. Mechanically seeded dislocations

For comparison, samples from the same batch (that has very low dislocation density as above) were particularly treated with an additional grinding and polishing step to purposely induce dislocations in the near-surface region. To this end, the samples were first ground using grinding paper (SiC particles, P2500), which can induce a large amount of surface dislocations with a density up to ~$10^{14}/m^2$ without microcracking (Supplementary **Fig. S2-3** and Supplementary **Movie S1**). Grinding can induce a very high density of dislocations as the microparticles on the grinding paper serve as stress concentrators for their nucleation. Further dislocation multiplication may occur during the grinding process. A spherical microparticle pressing into the sample surface can be considered as a spherical indenter. Related studies have been presented on $SrTiO_3$ (*25*), where profuse dislocation generation can be achieved, penetrating several micrometers underneath the sample. With the lateral shear force exerted during grinding, the generation of dislocations can be further facilitated. To further reduce the surface roughness and obtain a smooth surface for the following nano-/micropillar compression tests, the samples were subjected to additional steps of fine polishing with diamond paste for 30 seconds for each step, with a particle size of 6 μm, 3 μm, 1 μm, and 0.25 μm, respectively. Finally, a vibrational polishing for 30 min with OP-S solution (Struers, Copenhagen, Denmark) with an average particle size of 50 nm was performed. After this serial surface treatment, the final surface roughness is around 1~3 nm.

This procedure generates mechanically seeded dislocations in the near-surface region with a depth of about 2 μm (Supplementary **Fig. S2-3**). These dislocations belong to the room-temperature slip systems <110>{1-10} in SrTiO$_3$. In **Fig. 2-3** in the Main text, these dislocations are demonstrated to be mobile and serve as effective sources for dislocation multiplication during mechanical deformation.

## *S1.3. Micro-/Nanopillar compression: pillar preparation and mechanical testing*

### *A) Nanopillar compression*

Samples for in situ nanopillar compression were also fabricated by means of another FIB (Helios Nanolab 600i, FEI, Hillsboro, USA). Final dimensions with height of ~1200 nm and a width of ~700 nm (the thickness was about ~300 nm for optimized in situ recording of the evolution of dislocation structure) were machined and the sample was placed within a microelectromechanical systems (MEMS) chip. In situ nanopillar compression was performed using a double tilt heating/straining TEM holder (Bestron-INSTEMS-MT). The holder enables controllable compression of the sample and a range of ±25° α-tilt and ±25° β-tilt. The sample was tilted to a desired crystallographic zone axis for SAED (Selected Area Electron Diffraction) analysis. The compression unit was driven by a lead zirconate titanate (PZT) actuator under displacement control (smallest step size of 0.1 nm). The samples were aligned to a low-index zone axis (i.e., [001]) of the sample before in situ compression.

### *B) Micropillar compression*

A focused ion beam (FIB; equipment JEOL JSM 7600F) was used to prepare the micropillars. In order to reduce the FIB damage to the materials, a multi-step milling process was used with a beam current of 1000 pA and 500 pA during coarse milling, and 300 pA as final milling current. The same FIB milling parameters are applied for both types of micropillars with and without mechanically seeded dislocations. An overview of two representative pillars (one with and one without seeded dislocations) before the compression test is presented in Supplementary **Fig. S4**.

The micropillar compression tests were performed using a custom-made in situ load-controlled indentation system (*62*) (Prometheus, KLA instruments, Oak Ridge, USA) equipped with a diamond flat punch indenter (Synton-MDP, Nidau, Switzerland) with a diameter of ~10 μm for the micropillar compression tests. All tests were performed at room temperature inside a SEM (Vega 3, Tescan, Brno, Czech Republic) applying a pseudo-displacement-controlled loading protocol with a displacement rate of 10 nm/s, giving an equivalent strain rate of about $10^{-3}$/s considering the height of the micropillar is about 8 μm. The pillars were first deformed to the target strain (e.g., 10% and 25%) and then stopped for later cross-sectional check of the change in dislocation structure inside the pillar volume using FIB-fabricated TEM lamella. The stress-strain curves as well as the fracture strength (for pillars without seeded dislocations) and yield strength (for pillars with seeded dislocations) are summarized in Supplementary **Fig. S5** and **Fig. S6**.

### *C) Large spherical ball indentation and scratching tests*

Large spherical ball indentation (with an indenter tip diameter of 2.5 mm) as well as scratching tests were performed on a hardness tester (Karl-Frank GmbH, Weinheim-Birkenau, Germany), mounted with a Brinell indenter and a movable stage (PI Line M-6832V4, Physik Instrumente (PI) GmbH & Co. KG, Karlsruhe, Germany). A hardened steel spherical indenter (Habu Hauck Prüftechnik GmbH, Hochdorf-Assenheim, Germany) was used for testing. The sample was either first indented or scratched on the sample surface for one cycle (the 1$^{st}$ cycle indentation or scratching) to induce mechanically seeded dislocations (the moving speed of the indenter is 0.5 mm/s during the scratching test). Then either subsequent indentation or scratching test was performed for multiple cycles on the same spot after the 1$^{st}$ cycle, but now with newly induced seeded dislocations (see later **Figs. S7-8**). The indented or scratched area was then chemically etched to reveal the surface dislocation etch pits. For the scratch track, additional TEM examination of the dislocations within the scratch track, underneath the surface, was carried out (**Fig. S8**).

For estimating the density of existing dislocations in the deformed volume during spherical indentation or scratching (even before the 1$^{st}$ cycle testing), we take the case of a $R = 1.25$ mm (note the diameter of 2.5 mm was used being consistent). Upon indentation (right before horizontal moving for scratching), the indent imprint has a radius of about $r = 50$ μm. Consider a sample with a grown-in dislocation density of $10^{10}$/m$^2$ (see **Fig. S1C**), which corresponds to 0.01 dislocation/μm$^2$: then within the contact imprint (the area is $\pi r^2 = 7850$ μm$^2$ there will be around 80 existing dislocations forming the dislocations to be later activated (multiplied and moved) to accommodate the plasticity. In this case, plasticity is dominated by heterogeneous nucleation or activation of existing dislocations (multiplication and motion).

*S1.3. TEM sample preparation and dislocation characterization*

For TEM, thin lamella with a thickness of about 50 nm were prepared and thinned by a FIB (Helios Nanolab 600i, FEI, Hillsboro, USA). TEM samples were characterized by using a field emission TEM (Talos F200X G2) under 200 kV. For high-resolution scanning TEM (STEM) imaging and atomic energy dispersive X-ray spectroscopy (EDS), a double aberration-corrected microscope (FEI Titan Themis G2) was utilized at an acceleration voltage of 300 kV. High-resolution HAADF-STEM imaging was performed in a double aberration-corrected transmission electron microscope (FEI Titan Themis G2) operated at 300 kV. For HAADF imaging a probe semi-convergence angle of 17 mrad and inner and outer semi-collection angles ranging from 73 to 200 mrad were used. For ADF (annular dark-field) mode a probe semi-convergence angle of 17 mrad and inner and outer semi-collection angles ranging from 14 to 63 mrad were selected. For ABF imaging a probe semi-convergence angle of 17 mrad and inner and outer semi-collection angles ranging from 13 to 21 mrad were utilized.

Microstructures of STO including dislocations were characterized by a 200 kV-TEM with STEM mode. Tilt-series of ABF and ADF-STEM images were obtained using the Xplore3D software (FEI Company) taking dynamic focus into consideration. A high-angle tomography holder (Model 2020, Fischione, USA) was employed for the tilt-series acquisition with the maximum tilt angle of 60°, which is required to visualize the same set of dislocations in all the tilt-series images. In these tests, the [001] direction of the specimens (needle and lamella shapes) were aligned to the tilt axis of the specimen holder. We then used Inspect3D™ software (Thermo Fisher Scientific Inc.) to align the tilt-series images and reconstruct the 3D datasets by the simultaneous iterative reconstruction technique (SIRT) with 40 times iterations. All the tilt-series STEM images were used for 3D reconstruction, and Amira-Avizo™ software was then used for 3D display. All crystallographic analyses were assisted by contrasting the experimental SAED patterns to computed electron diffraction patterns and the corresponding stereographic pole figure analysis was conducted by utilizing electron microscopy simulation software (Crystalmaker, with the crystallographic information, CIF file) under an assumption of a 50 nm TEM foil/lamella thickness and an acceleration voltage of 300 kV. The dislocation line vector and Burgers vector analyses are summarized in **Figs. S9-11** to help determine which type of dislocations are present due to mechanical seeding.

**Supplementary Text:**

Referenced literature for the pillar compression of different materials for **Fig. 4A** in the main text. Note the materials are single crystals unless specified. The numbers in the parentheses indicate the literature number in the main manuscript file.

- SrTiO$_3$: (*44*)
- MgO: (*70*) & (*71*)
- ZnO: (*72*)
- β-Ga$_2$O$_3$: (*73*)
- MgAl$_2$O$_4$: (*74*)
- BaTiO$_3$: (*45*)
- ZrO$_2$: polycrystalline (*75*)
- TiO$_2$: polycrystalline (*19*)
- Si: (*41*) & (*76*)
- GaAs: (*77*) & (*43*)
- AlN: (*78*)

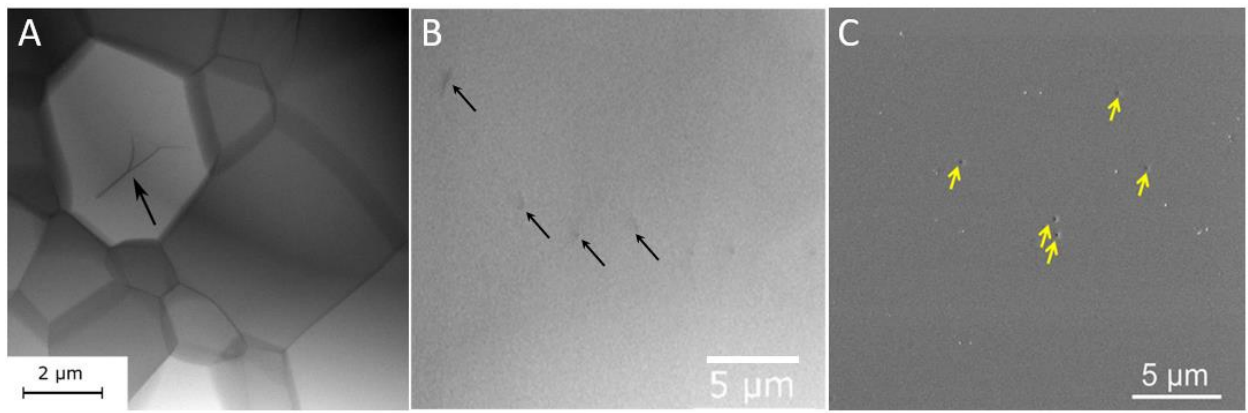

**Fig.S1.** (A) high-voltage TEM of a polycrystalline SrTiO$_3$ sintered at ~1425 $^o$C; (B) high-voltage TEM of a single-crystal TiO$_2$ grown from Verneuil technique; (C) SEM image of a single-crystal SrTiO$_3$ surface after chemical etching. The pre-existing dislocations are indicated with black or yellow arrows. All three cases feature extremely low dislocation density (lower than $10^{10}$/m$^2$) in the as-prepared samples. **Fig. S1A** adopted from Ref. (*20*) and **Fig. S1B** from Ref. (*46*) with permission.

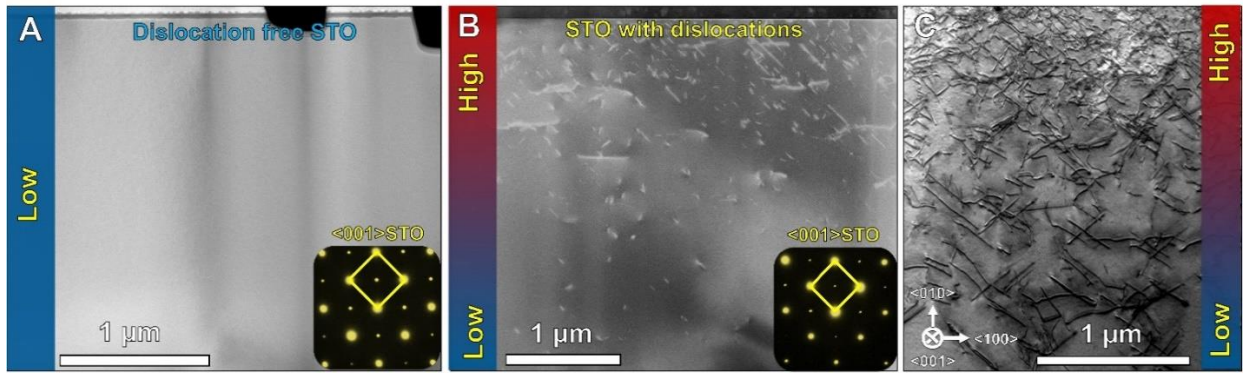

**Fig. S2.** (A) Annular dark-field scanning transmission electron microscope (ADF-STEM) image provides the pristine dislocation-free SrTiO$_3$. (B) LAADF (low-angle annular dark-field)-STEM image displays the mechanically seeded dislocations in the near surface region of (001) SrTiO$_3$ crystal. A gradient of the dislocation distribution is observed along the depth. (C) Annular bright field (ABF)-STEM image of the mechanically seeded dislocations from a different viewing angle as in (B), displaying more line features. Later the mechanical tests were performed in dislocation-free and dislocation-rich regions.

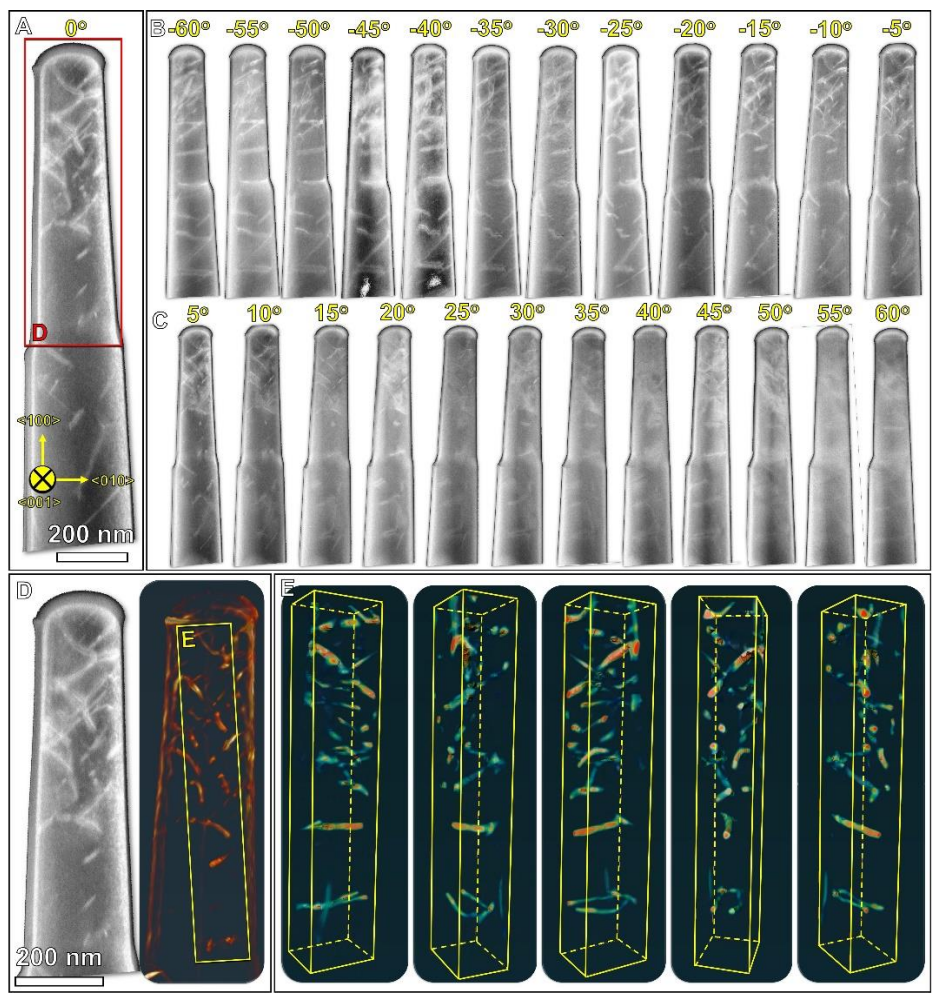

**Fig. S3.** 3D reconstruction of SrTiO$_3$ tip with pre-induced dislocations. (A-B) ADF-STEM image of sample needle along <001> zone axis. (B) Series ADF-STEM images tilted along <100> axis from -60° to 60°. (C) Enlarged area from (A) reveals the dislocation networks and the corresponding 3D image. (D) Cropped region from (marked in **Fig. S3A** with red box as D) displays the series 3D reconstructed dislocations along different viewing directions.

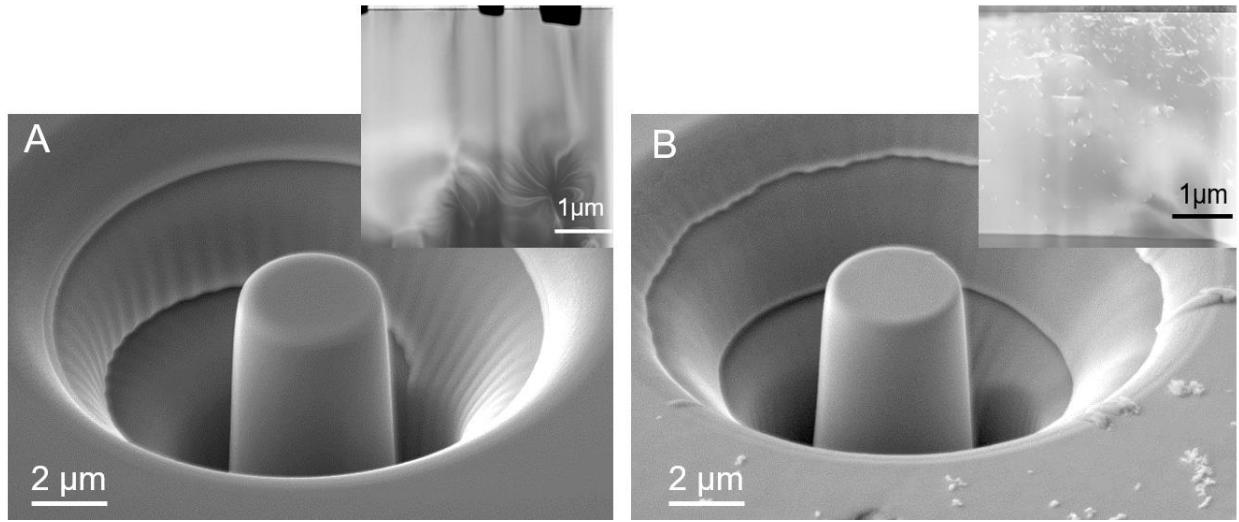

**Fig. S4**: SEM images of the micropillars before compression tests: (A) pillar prepared from pristine sample, with the insert TEM image revealing the complete absence of dislocations in the samples; (B) pillar prepared from samples with mechanically seeded dislocations, with the insert TEM image displaying abundant dislocations (white contrast) similar as in **Fig. S2B**.

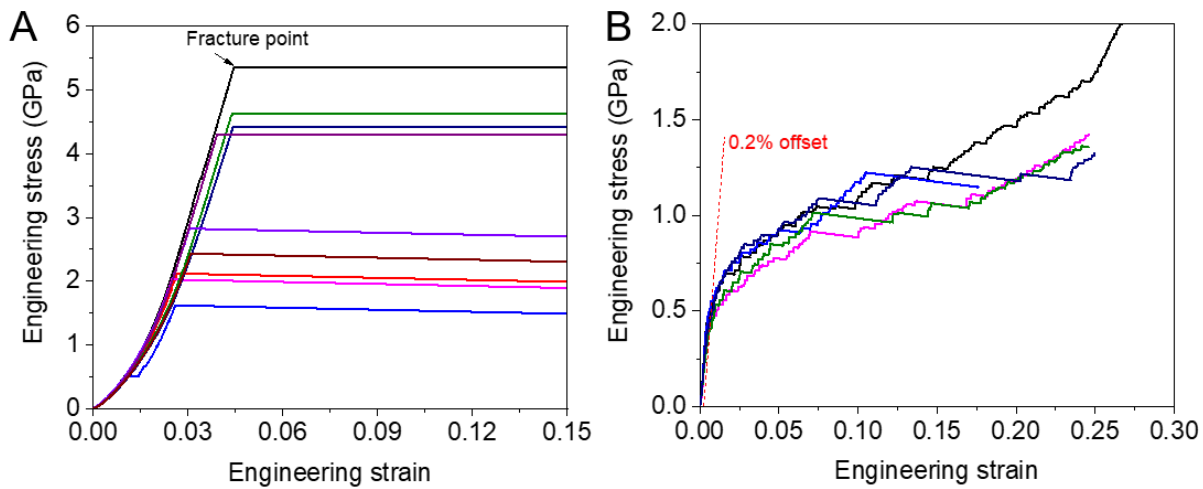

**Fig. S5:** Engineering stress-strain curves for the micropillars without (A) and with (B) pre-engineered seeded dislocations. The reproducibility is ensured by testing multiple pillars. Note that in the case of the pillars without seeded dislocations the scatter is much larger. The yield strength for the pillars with seeded dislocations was determined using the 0.2% strain offset.

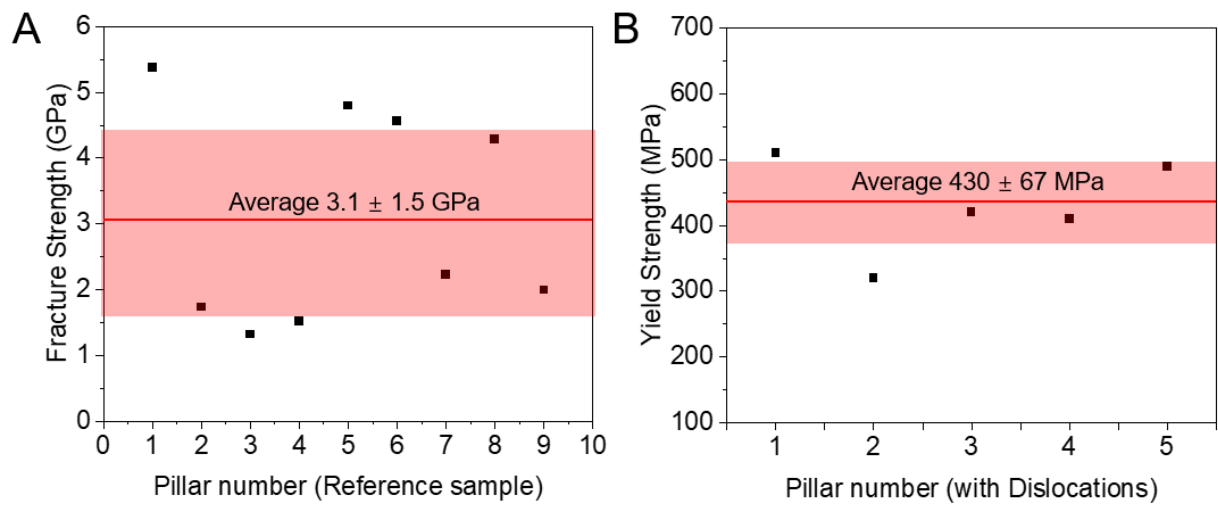

**Fig. S6:** Statistical distribution of the (A) fracture strength for micropillars without seeded dislocations; (B) yield strength for micropillars with pre-engineered seeded dislocations. The average values (red lines) and the standard deviation (red shaded zones) are indicated for both cases.

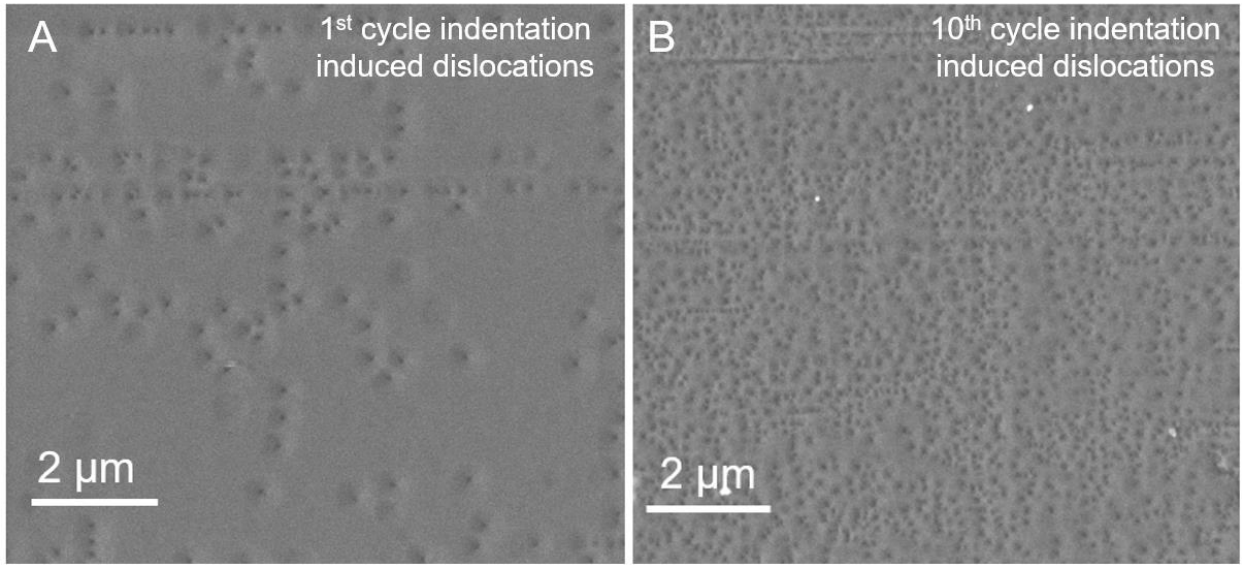

**Fig. S7**. 1st cycle spherical indentation: Induced dislocations (A) serve as mechanically seeded dislocations for the subsequent dislocation multiplication (B) after 10th cycle on similar location as in (A).

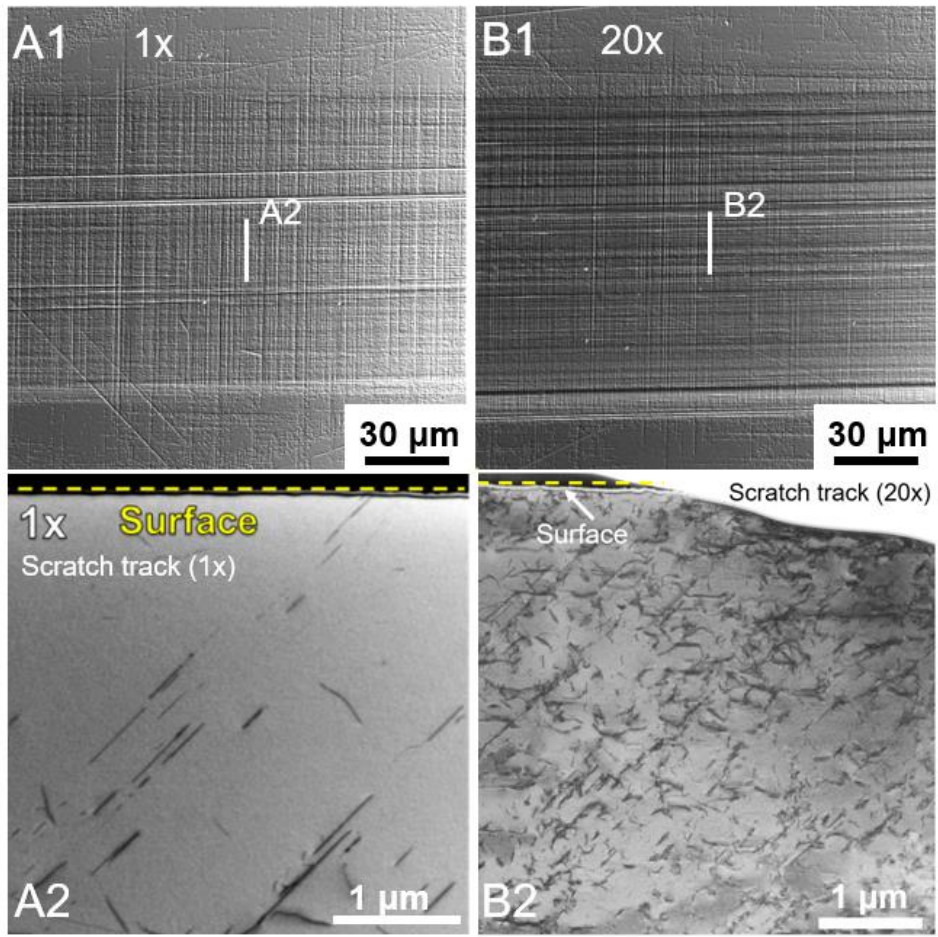

**Fig. S8.** Dislocation multiplication during multiple scratch tests on (001) SrTiO$_3$ at room temperature using a large spherical ball indenter *(63)*: Pre-engineered dislocations are achieved by a first-round scratching: (A1) laser microscope image of the surface dislocation etch pits after chemical etching; (A2) TEM images of the dislocations (dark lines) below the scratched surface. Multiple scratch deformation up to 20 cycles (20x) in the same scratch track with pre-engineered dislocations after the first-round scratching: (B1) laser microscope image of the surface dislocation etch pits after chemical etching; (B2) TEM images of the dislocations (dark lines) below the scratched surface. Profuse dislocations are generated due to the pre-engineered seeded dislocations, and clear evidence of dislocation multiplication was observed, consistent with the in situ observation of the nanopillar compression in **Fig. 2** in the main text, although the stress states under uniaxial compression and scratching differ. Sub-figures A1, B1, B2 adapted from Ref. *(63)* (under Creative Commons Attribution License, CC-BY).

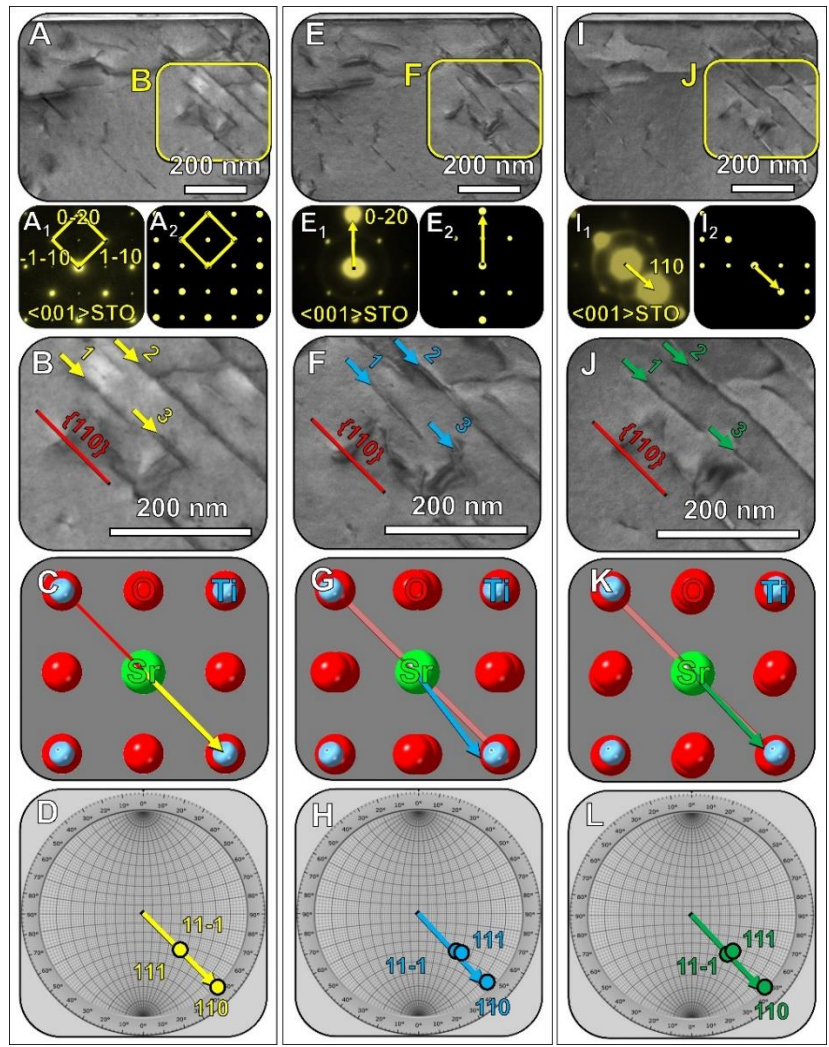

**Fig. S9.** Dislocation analysis of the mechanically seeded dislocations: (A) and (B) dislocation lines along <001> zone axis. (C) Schematic illustrating the traces of the dislocations and {110} slip plane of SrTiO$_3$ along <001> zone axis. (D) Pole figure displays the trace analysis of the linear dislocations along <100> zone axis. (E) and (F) Tilting under g = 0-20 highlights the change of dislocation traces. (G) Schematic detailing the change of dislocation traces and {110} slip plane of SrTiO$_3$ under g = 0-20. (H) Pole figure displays the trace analysis of the linear dislocations under g = 0-20. (I) and (J) Tilting under g = 110 illustrates the change of dislocation traces. (K) Schematic highlighting the change of traces of the dislocation segment and {110} slip trace of SrTiO$_3$ under g = 110. (L) Pole figure depicts the trace analysis of the dislocation segment under g = 110. The line vector of dislocations (labelled with 1,2,3) is 11-1 according to the trace analysis, indicating mixed type dislocations.

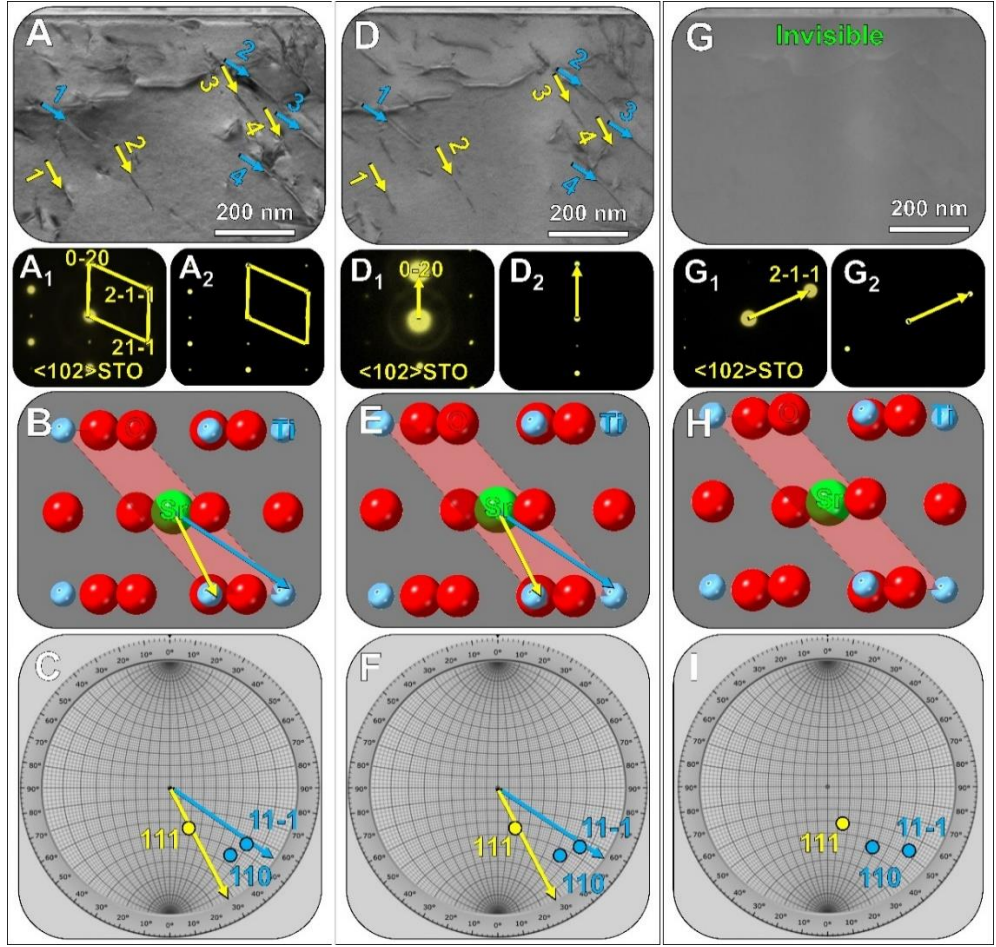

**Fig. S10.** Dislocation analysis of the mechanically seeded dislocations SrTiO$_3$. (A) Linear dislocations along <102> zone axis. (B) Schematic illustrating the trace of the linear dislocation and {110} slip direction. (C) Pole figure features the traces and line vectors of the linear dislocation. (D) Visible linear dislocations under g = 0-20. (E) Schematic diagram demonstrates the change of traces of linear dislocations under g = 0-20. (F) Pole figure highlights the traces and line vector of the linear dislocations under g = 0-20. (G) Dislocations invisible under g = 2-1-1. (H) Schematic displaying the slip plane. (I) Pole figure illustrating the possible line vector under g = 2-1-1. According to the trace analysis, the line vector of 11-1 is identified for dislocations labelled with 1-4 in blue arrows, and the line vector for dislocations labelled with 1-4 in yellow arrows is 111. According to the dislocation invisible criterion, the Burgers vector yields 01-1. Hence, the dislocations in the blue arrows are 35.26° mixed type (in agreement with Ref. (*49*)), and the dislocations in the yellow arrows yield edge type.

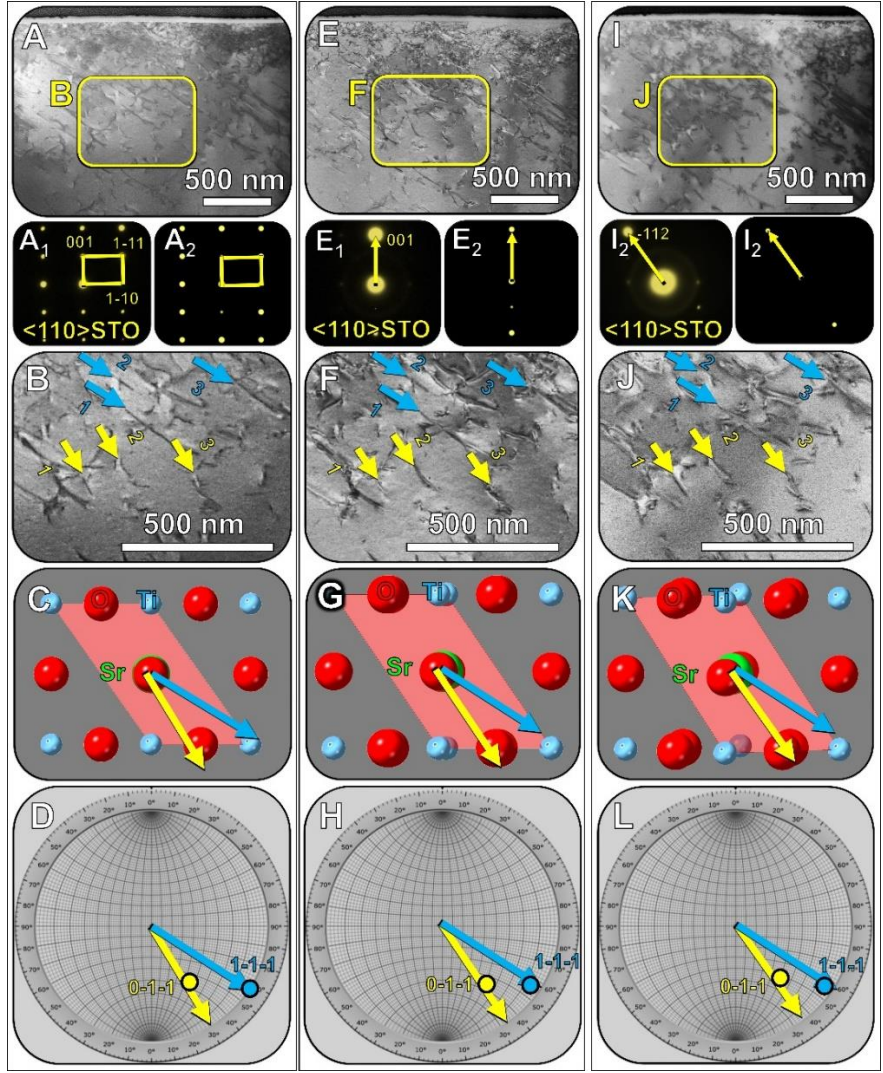

**Fig. S11.** Dislocation analysis of the mechanically seeded dislocations in SrTiO$_3$. (A) and (B) visible dislocations along <110> zone axis. (C) Schematic diagram details the traces of the dislocations and slip direction. (D) Pole figure illustrates the traces and line vectors of the dislocations. (E) and (F) Visible dislocations under g = 001; (G) Schematic diagram highlights the change of traces of linear dislocations under g = 001. (H) Pole figure provides the traces and line vectors of linear dislocation under g = 001. (I) and (J) Visible dislocations under g = -112. (K) Schematic displaying the slip plane. (L) Pole figure illustrates the possible line vectors under g = -112. According to the trace analysis, the line vector of dislocations 1-3 in blue arrow yields 1-1-1, and the line vector yields 0-1-1 for dislocation 1-3 in yellow arrows.

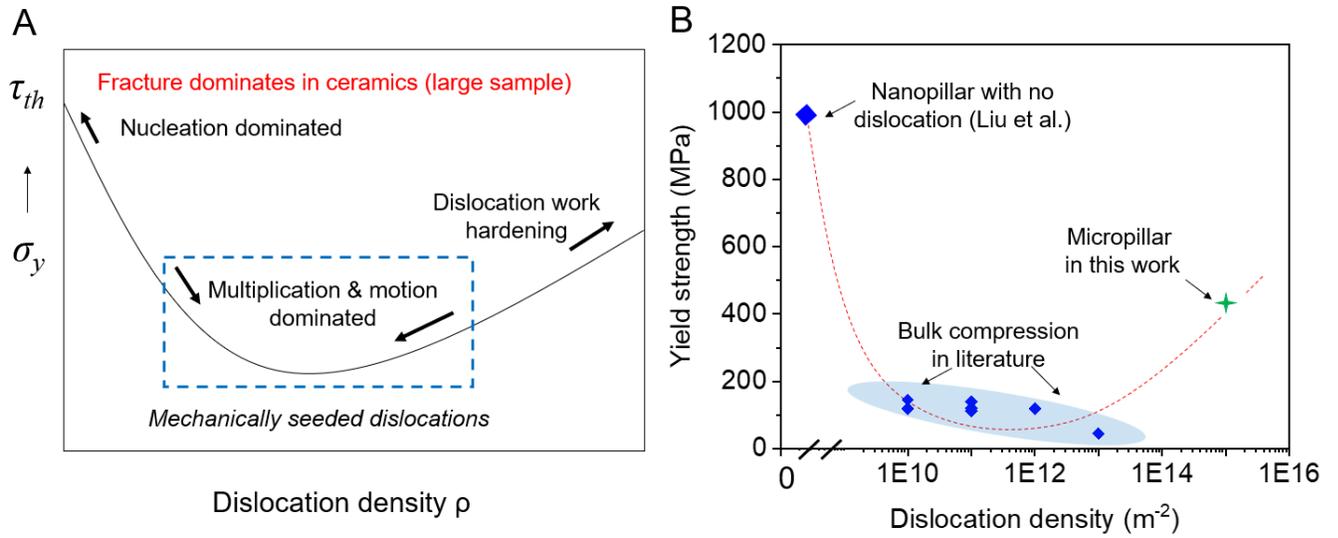

**Fig. S12.** (A) Size effect in metals, schematic modified based on Ref. *(53)*; (B) Yield strength as function of dislocation density in single-crystal SrTiO$_3$ with data extracted from literature: Liu et al. (*44*), bulk compression data please see later Supplementary **Table S1**.

**Table S1.** Summary of yield strength values for single-crystal SrTiO$_3$ along [001] direction in compression tests performed at room temperature (RT) (data utilized in Supplementary **Fig. S12B**)

| | Yield strength (MPa) | Strain rate (s$^{-1}$) | Sample geometry (mm$^3$) | Pre-existing dislocation density ρ (m$^{-2}$) | Dislocation spacing 1/√ρ (mm) | Pre-existing dislocation number | Maximum RT plastic strain |
|---|---|---|---|---|---|---|---|
| Stich et al. (*64*) | 145 | 10$^{-4}$ | ~ 2 x 2 x 4 | ~ 10$^{10}$ | ~1 x10$^{-2}$ | ~4x10$^5$ | NA |
| Brunner et al. (*54*) | 120 | 10$^{-4}$ | 2.5 x 2.5 x 6 | ~8 x 10$^{10}$ | ~3 x10$^{-3}$ | ~6x10$^6$ | ~7% |
| Brunner (*65*) | 120 | 2.1 x 10$^{-4}$ | 2.5 x 2.5 x 6 | ~8 x 10$^{10}$ | ~3 x10$^{-3}$ | ~6x10$^6$ | ~7% |
| Taeri et al. (*29*) | 140 | 10$^{-4}$ | 2.5 x 2.5 x 6 | ~8 x 10$^{10}$ | ~3 x10$^{-3}$ | ~6x10$^6$ | ~6% |
| Yang et al. (*66*) | 45 | 10$^{-4}$ | 3 x 3 x 6 | ~4.8 x 10$^{13}$ | ~1.4 x10$^{-4}$ | ~4x10$^9$ | ~19% |
| Patterson et al. (*67*) | 118 | 10$^{-4}$ | 4 x 4 x 8 | ~ 10$^{12}$ | ~1 x10$^{-3}$ | ~1x10$^8$ | NA |
| | 120 | 10$^{-5}$ | | | | | NA |
| Nakamura et al. (*68*) | 119 | 10$^{-5}$ | 3 x 3 x 7.5 | ~ 10$^{10}$ | ~1 x10$^{-2}$ | ~1.1x10$^6$ | 7.4% |
| | 112 | 10$^{-5}$ | 3 x 3 x 7.5 | ~ 10$^{11}$ | ~3 x10$^{-3}$ | ~1.1x10$^7$ | 13.6% |
| Li et al. (*69*) | 120 | 10$^{-4}$ | 2 x 2 x 4 | NA | NA | NA | NA |
| Liu et al. (*44*) | 1.6 GPa (brittle cracking) | 10$^{-4}$ | Nanopillar (150 nm < diameter< 260 nm) | Dislocation-free (based on TEM image) | >3 x10$^{-4}$ | 0 | NA |
| Liu et al. (*44*) | 1 GPa (plastic yield) | 10$^{-4}$ | Nanopillar ~150 nm in diameter | Dislocation-free (based on TEM image) | >3 x10$^{-4}$ | 0 | ~15% |
| This work | ~3.1 GPa (brittle cracking) | 10$^{-4}$ | Micropillar (~4 μm) | ~ 10$^{10}$ | ~1 x10$^{-2}$ | ~1 | NA |
| This work | ~430 MPa (plastic yield) | 10$^{-4}$ | Micropillar (~4 μm) | ~ 10$^{15}$ | ~1 x10$^{-4}$ | ~1x10$^4$ | >32% |

*__Note 1__: The rows highlighted in red were not used: Ref. (54) was used (Ref. (65) not) due to the fact that this is likely the same batch of data from the same group of authors. Ref. (69) is not used for plotting as the dislocation density in the sample was unclear. Also, the data from Liu et al. in Ref. (44) was not used due to the missing information of the dislocation density. The rows highlighted in green are from this current work.*

*__Note 2:__ The number of dislocations in the deformation volume was estimated by using the dislocation density multiplied by the area of the deformation volume. For instance, if a sample has a geometry of 2x2x4 mm3, then the surface area would be: 2x2x2 (2 top/bottom surfaces) + 2x4x4 (4 side surfaces) = 40 mm$^2$ = 4x10$^{-5}$ m$^2$. With a density of 10$^{10}$/m$^2$, an estimated number of 10$^{10}$/m$^2$ x 4x10$^{-5}$ m$^2$ = 4x10$^5$ in the whole sample is obtained.*

*__Note 3:__ the data by Patterson et al. in Ref. (67) suggest that the yield strength is negligibly affected by the strain rate with only one order of magnitude change.*